\documentclass[prl,twocolumn,superscriptaddress]{revtex4}

\usepackage{float}
\usepackage{graphicx}
\usepackage{color}
\usepackage{amssymb}
\usepackage{amsmath}
\usepackage{accents}
\usepackage{hyperref}
\usepackage[normalem]{ulem}

\newcommand{\kon}{ k_{\rm on}}
\newcommand{\koff}{k_{\rm off}}

\newcommand{\be}{\begin{equation}}
\newcommand{\ee}{\end{equation}}
\newcommand{\bea}{\begin{eqnarray}}
\newcommand{\eea}{\end{eqnarray}}

\newcommand{\meanVa}{\langle V_a \rangle}

\newcommand{\meanKa}{\langle K_a\rangle}

\newcommand{\fcoll}{f_{\text{coll}}}

\newcommand*{\helv}{\fontfamily{phv}\selectfont\bfseries}

\begin{document}
\title{Modelling Protein Target-Search in Human Chromosomes}
\date{\today}

\author{Markus Nyberg}
\affiliation{Integrated Science Lab, Department of Physics, Ume\r{a} University, SE-901 87 Ume\r{a}, Sweden}
\author{Tobias Ambj\"ornsson}
\affiliation{Department of Astronomy and Theoretical Physics, Lund University, SE-223 62 Lund, Sweden}
\author{Per Stenberg}
\affiliation{Department of Ecology and Environmental Science, Ume\r{a} University, SE-901 87 Ume\r{a}, Sweden}
\author{Ludvig Lizana}
\email{ludvig.lizana@umu.se}
\affiliation{Integrated Science Lab, Department of Physics, Ume\r{a} University, SE-901 87 Ume\r{a}, Sweden}

\begin{abstract}
Several processes in the cell, such as gene regulation, start when key proteins recognise and bind to short DNA sequences. 
However, as these sequences can be hundreds of million times shorter than the genome, they are hard to find by simple diffusion: diffusion-limited association rates may underestimate {\it in vitro} measurements up to several orders of magnitude. Moreover, the rates increase if the DNA is coiled rather than straight.
Here we model how this works {\it in vivo} in mammalian cells. We use chromatin-chromatin contact data from state-of-the-art Hi-C experiments to map the protein target-search onto a network problem. The nodes represent a DNA segment and the weight of the links is proportional to measured contact probabilities. We then put forward a master equation for the density of  searching protein that allows us to calculate the association rates across the  genome analytically.  For segments where the rates are high,  we find that they are enriched with active genes and have high RNA expression levels. This paper suggests that the DNA's 3D conformation is important for protein search times {\it in vivo} and offers a method to interpret protein-binding profiles in eukaryotes that cannot be explained by the DNA sequence itself.

\end{abstract}

\maketitle

Several processes in the cell nucleus start when proteins bind to specific DNA sequences. For example, transcription factors that regulate genes and the CRISPR/CAS9 complex that edits DNA \cite{farnham2009insights,globyte2019crispr}. Because target sequences are much shorter than the genome --  a few base pairs compared to billions in humans -- these proteins face a needle-in-a-haystack problem.

Despite the large number of potential targets, measured search times are shorter than theoretical estimates. The Lac repressor in {\it E. coli}  needs 1-5 min  to find its designated site \cite{hammar2012lac} which is  twice as fast as a three-dimensional (3D) search by diffusion inside the bacterium's volume ($\approx $2-11 min) \footnote{Diffusion-limited search time, $\tau = 4\pi a D/V$ where $D=0.1-0.5$, $\mu$m$^2$/s, $a=5$ bp (=1.3 nm), and $V = 1 \mu$m$^3$. These values give $\tau=2.1 - 10.7$ min.}. 
Also, diffusion-limited association rates -- Smoluchowski's rate -- may underestimate {\it in vitro}  measurements by one to two orders of magnitude \cite{riggs1970lac}.  These examples suggest that some proteins search by other mechanisms than  simple diffusion.

One  mechanism that speeds up diffusive search is offered by the Facilitated-diffusion model \cite{von1989facilitated}. In this model, the proteins alternate between 3D diffusion and 1D diffusion along the DNA. This decreases the search time because the proteins may take shortcuts to a linearly distant DNA segment through the surrounding bulk. Although criticized \cite{kolomeisky2011physics,mirny2009protein},  the  model is widely accepted after experiments in bacteria \cite{elf2007probing,hammar2012lac} and  {\it in vitro}~\cite{van2008dna}. 

Another important aspect in target finding is  rebinding. This is because proteins  likely bind to a DNA segment that is close by in 3D rather than far away. Several modelling studies examined this aspect and found that search times change with DNA conformation \cite{amitai2018chromatin, van2008dna, bauer2013vivo, hu2006proteins,lomholt2009facilitated,lomholt2005optimal,mirny2009protein}. However,  because these studies treat the DNA as a simple polymer the results cannot  be generalized beyond bacteria to eukaryotes that have longer DNA and more complex 3D  organization.

\begin{figure*}
\includegraphics[width=\textwidth]{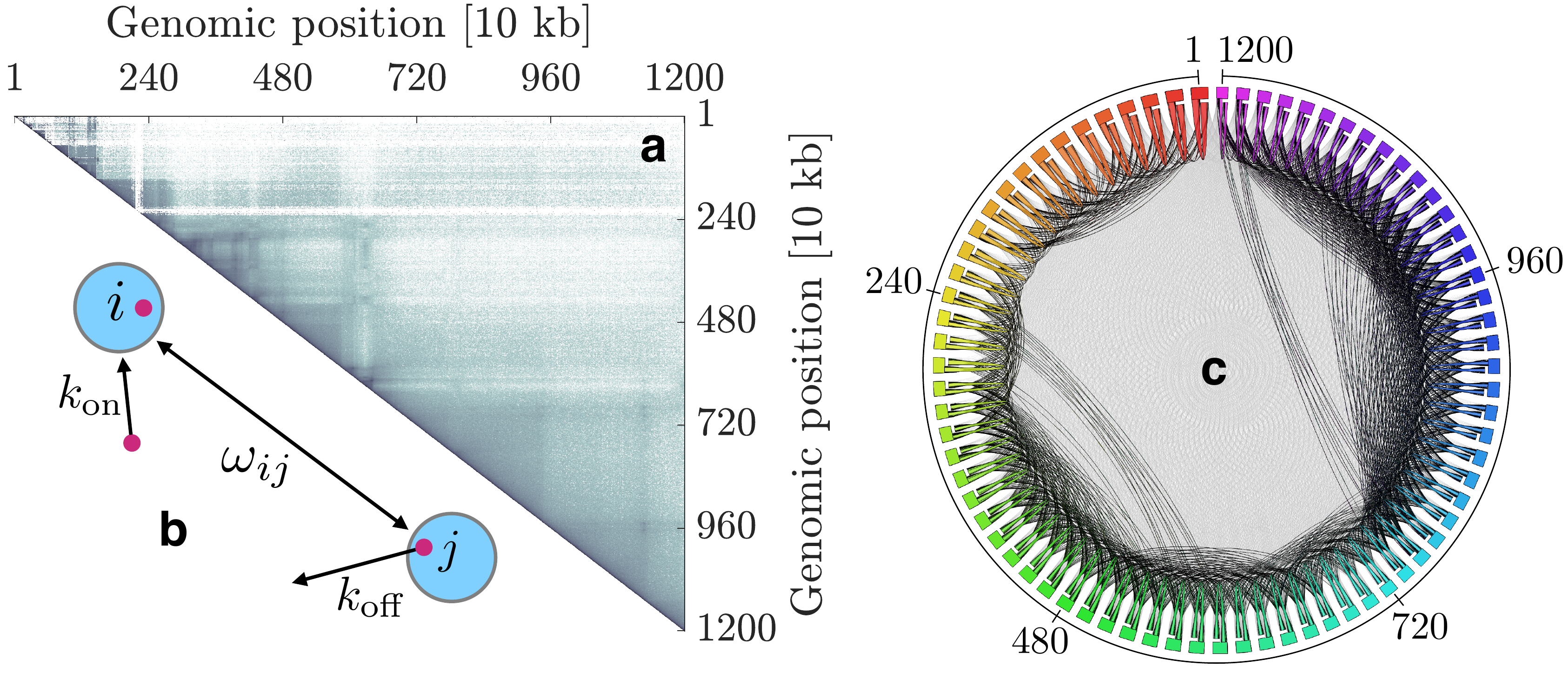}
\caption{Modelling protein search on DNA as search on a weighted network. 
(a)  Hi-C map: the number of pairwise  physical contacts between 10 kilo-basepair (kb) DNA fragments in a part of human chromosome 21. Dark pixels indicate many contacts.
(b) Schematic representation of the model. The key parameters are: jump rates between nodes $i$ and $j$ ($\omega_{ij}$), unbinding rate to the bulk ($\koff$), and rebinding rate ($\kon$). Red circles represent searching proteins.
(c) Coarsed network representation of the Hi-C map in (a) where each node represents a 160 kb fragment. The link weights $v_{ij}$ are proportional to the number of Hi-C contacts. We assume that $\omega_{ij}\propto v_{ij}$. Node numbering refers to positions along the DNA. 
\label{fig:hic+network}}
\end{figure*}

The most widely used experimental method to study 3D genome organization is Hi-C \cite{lieberman2009comprehensive,kong2019deciphering}. The Hi-C method cross-links close by DNA fragments inside the nucleus and gives a genome-wide  map of the number of contacts between fragments pairs  (Fig. \ref{fig:hic+network}a) \cite{rao20143d}. Mamalian Hi-C maps have several interesting features -- some that are evolutionary conserved \cite{krefting2018evolutionary}. For example,  the  block--like structure along the diagonal represents densely connected 3D domains. The locations of these domains correlate with protein binding sites, active genes, and chromatin states \cite{dixon2012topological,jost2014modeling,lee2019mapping}.  And, beyond the domains, the average contact probability decays as a power-law \footnote{One over distance in humans ($>7\times 10^5$ base pairs \cite{lieberman2009comprehensive}}. 

Hi-C is the state-of-the-art Chromosome Conformation Capture method that estimates the chromatin contact probabilities across the genome. But it does not provide chromatin's 3D structure. Going from the contact map to a computer-generated 3D structure is difficult \cite{di2017novo,serra2015restraint}. 

Because the chromatin's spatial organization is so complex, there are few attempts to model protein search in eukaryotes. One exception \cite{smrek2015facilitated},  represents chromatin as a crumpled polymer globule that reproduces the average looping probabilities in the human genome. However, the crumpled globule lacks the Hi-C maps' domain structure.

Here we model target search in eukaryotes without relying on chromatin's explicit 3D structure. Instead, we represent the DNA as a network in which the nodes are DNA segments and the link weights are proportional to the contact probabilities measured in Hi-C. Then we put forward a master equation for the density  of  proteins on the network over time that allows us to calculate the  association rate -- the inverse  mean-first passage time --  to all nodes analytically. Correlating these rates with RNA expression data in humans, we find that easy-to-find genomic regions are enriched with active genes and have high RNA expression.



\section{The model}

We model the proteins' search on chromatin as non-interacting particles that move  between nodes in a weighted network that represent physically connected chromatin segments (Fig. \ref{fig:hic+network}). 

The model has three important parameters. First, the jump rate $\omega_{ij}$ between nodes $i$ and $j$ ($i,j=1,\ldots,N)$. We assume that $\omega_{ij}$ is proportional to the number of contacts between segments $i$ and $j$, $v_{ij}$, measured in Hi-C: $\omega_{ij} = \fcoll v_{ij}$; $\fcoll$ -- a free parameter in the model --  is the collision frequency that leads to a successful jump. Second, the unbinding rate $\koff$ to the surrounding bulk. And third, the rebinding rate $\bar{k}_{\rm on}$ to a randomly chosen node. We assume that the protein bulk concentration $n_{\rm bulk}$ is constant and therefore use $\kon = \bar{k}_{\rm on} n_{\rm bulk}$;  $\kon$ and $\koff$ therefore have the same unit: time$^{-1}$. 

In terms of these parameters, we formulate a master equation for the protein number in node $i$ at time $t$, $n_i(t)$:
%
\be\label{eq:master+dna}
\frac{d n_i(t)}{d t}=\sum_{j=1}^N\omega_{ij}n_j(t)-\koff n_i(t)+\kon
\ee
The first term represents diffusion on the network -- we put $\omega_{jj}=-\sum_{i\neq j}\omega_{ij}$ -- and the two remaining terms describe the exchange with the surrounding bulk.

As in \cite{lomholt2005optimal}, we assume that the protein density is initially uniform $\rho_0=\kon/\koff$ except for the target $i=a$  which always is zero $n_a(t)=0$, that is $n_i(0)=\rho_0(1-\delta_{ia})$.

In terms of the eigenvalues $\lambda_j$ and eigenvectors $V_{ij}$ of $\omega_{ij}$, the solution to Eq. \eqref{eq:master+dna} is
\begin{eqnarray}\label{eq:ni(t)}
n_i(t) &=\sum_{j=1}^N V_{ij} \bigg\{ \frac{\kon^j}{k_{\rm off}-\lambda_j}\left[ 1-e^{-(k_{\rm off}-\lambda_j)t} \right] 
\nonumber\\
& +e^{-(k_{\rm off}-\lambda_j)t}\rho_0\sum_{l\neq a}V^{-1}_{jl} \bigg \}.
\end{eqnarray}
where $\kon^i=\kon\sum_{j=1}^NV^{-1}_{ij}$. This equation is key to calculate the association rate to the target node $i=a$.

\begin{figure*}
\includegraphics[width = \textwidth]{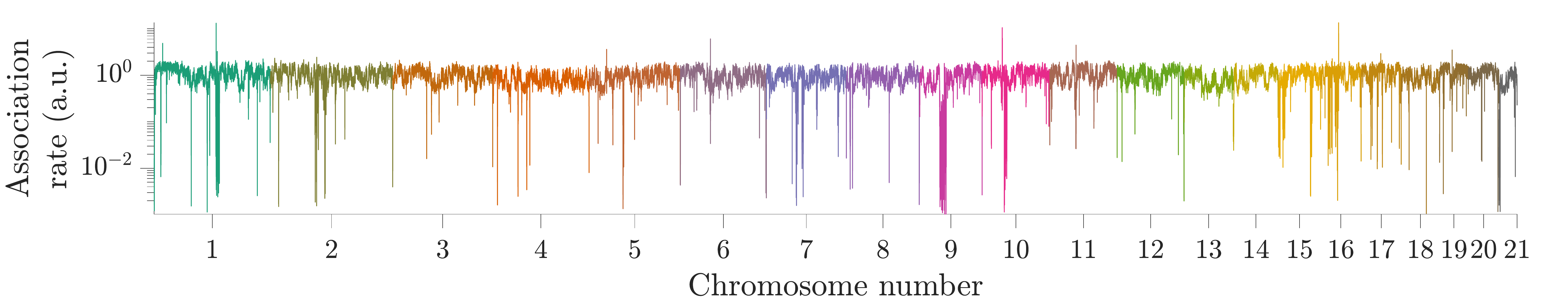}
\caption{Predicted genome-wide association rates $K_a$ for chromosomes 1-21. $K_a$ vary by several orders of magnitude along the chromosomes but the 95\% confidence interval is a few percent of the mean $\meanKa=1$ ($K_a = 1.0\pm0.0027$). We calculated $K_a$ from Eq. \eqref{eq:fast+Ka} ($\koff \gg 1)$ using the parameters $\koff = 0.002$, $\kon = 0.001$ and $\rho_0=0.5$.}
\label{fig:Ka_vs_chr}
\end{figure*}

\section{Protein association rates}
To calculate the association rate $K_a$, we use that it is one over the mean first arrival time:  $K_a=\tau_a^{-1}$. To obtain $\tau_a$, we proceed as in \cite{lomholt2005optimal}. First we calculate the total number of particles that arrived to the target up to time $t$: $J_a(t)=\int_0^tj_a(t')dt'$ where 
\be
j_a(t)=\sum_{i}\omega_{ai}n_i(t)+\kon
\ee
is the particle flux. Second, if $N_p$ is the initial protein number on the network, and if $\kon=\koff=0$, then $J_a(t)/N_p$ is the probability that a single protein have reached the target up to time $t$. The probability that the target has not been reached  -- the target's survival probability  -- is therefore $S_a(t) = (1-J_a(t)/N_p)^{N_p}$ and  $\tau_a=\int_0^{\infty}S_a(t)dt$.
%
%
Generalizing this argument for $\kon>0$, the number of proteins $N_p\to\infty$ and therefore $S_a(t)\simeq\exp\left(-J_a(t)\right)$ \cite{sokolov2005first}.  Note that if $\kon=0$ and $\koff>0$ there is a chance that all proteins end up in the bulk before ever reaching the target causing $\tau_a$ to diverge.

Finally, using $n_i(t)$ from Eq. \eqref{eq:ni(t)}, we may calculate $K_a$ exactly as:
\be\label{eq:Ka}
K_a = \frac 1 {\int_0^\infty \exp\left(-J_a(t)\right)dt},
\ee
%
%
\be\label{eq:Ja}
J_a(t) = \kon t + \sum_{i\neq a}\omega_{ai}\int_0^{t} n_i(t') dt',
\ee
%

\subsection{Limiting cases for $K_a$}
Depending on the unbinding rate $\koff$, $K_a$ has three regimes.

{\it (i) Small $\koff$.} In this regime most particles find the target before they unbind. This leaves the initial density $\rho_0$ approximately unchanged, $n_i(t)\simeq \rho_0$. Using this approximation in Eq.~\eqref{eq:Ja} leads to  $J_a(t)\simeq \bar J_a t$, where $\bar J_a = \kon + \rho_0 \sum_{i\neq a}^N\omega_{ai}\equiv \kon + \rho_0W_a$, and thus $K_a \simeq \bar J_a$.

{\it (ii) Large $\koff$.} Here, the particles unbind and rebind many times before finding the target. The protein density is therefore approximately in steady-state $\bar{\rho}_a=\kon \times[\koff+W_a/(N-1)]^{-1}$ \cite{suppl}. Using $n_i(t)\simeq\bar \rho_a$ and proceeding as in {\it (i)} leads to $K_a \simeq \kon+\bar \rho_a W_a$.

These regimes simplify to $\koff \gg 1$ and $\koff\ll 1$ if we choose $\fcoll$ so that the genome-wide averaged $K_a$, $\langle K_a \rangle$, is unity \cite{suppl}; We define the average as $\langle X\rangle = (1/N_G) \sum_{i=1}^{N_G}X_i$, where $N_G\gg N$ is the  number of nodes for all chromosomes 

After rescaling, we have
\be \label{eq:fast+Ka}
K_a \simeq \kon +  \gamma_a V_a,
\ee
in which $V_a=\sum_{i\neq a}^N v_{ai}$ is the node strength and
\be
\gamma_a = \frac{1 -\kon}{\meanVa} = \gamma_0,  \ \ \ \koff\ll 1
\ee
\be
\gamma_a = \frac {\kon \gamma_0}{\kon+V_a\gamma_0/(N-1)},  \ \ \ \koff\gg 1
\ee
%
Equation \eqref{eq:fast+Ka} covers a broad range of $\koff$ \cite{suppl}, it is easy to implement, and computationally cheaper than  Eqs. \eqref{eq:Ka} and \eqref{eq:Ja}. 

{\it (iii) Intermediate $\koff$.} When $\koff\sim 1$, we cannot use Eq. \eqref{eq:fast+Ka}. Instead we must use the exact expressions \eqref{eq:Ka} and \eqref{eq:Ja}. In \cite{suppl} we also treat the case $\kon=\koff=0$.

\subsection{Protein association rates depend on chromatin's 3D organization}

Equation \eqref{eq:fast+Ka} suggests that the association rates change with chromatin's 3D structure because $K_a$ depends on the node strength $V_a$. To quantify by how much, we used Hi-C data (40 kb resolution) and calculated $K_a$ for chromosomes 1-21 (Fig. \ref{fig:Ka_vs_chr}). We found that $K_a$ varies by several orders of magnitude relative to the genome-wide average $\langle K_a\rangle =1$. Most $K_a$ values, however, only deviate  by a few percent from the mean: $K_a = 1 \pm 0.0027$ (95\% confidence interval). Figure \ref{fig:Ka_vs_chr} shows the case when $\koff \ll 1$; $\koff \gg 1$ has qualitatively the same behavior~\cite{suppl}.

Equation \eqref{eq:fast+Ka} also suggests that chromatin's 3D structure becomes less important as the unbinding rate $\kon$ grows -- for example by increasing the bulk concentration of particles as $\kon\propto n_{\rm bulk}$. We see this in the data for small $V_a$  where $K_a \simeq \kon$  \cite{suppl}. We interpret this as if the particles reach the target mostly from the bulk.  In the other limit, where $\kon$ is small,  we see that $K_a\propto V_a$. This means that most particles find the target via jumps on the network and that the 3D structure is important. 



\begin{figure}
    \centering
    \includegraphics[width=\columnwidth]{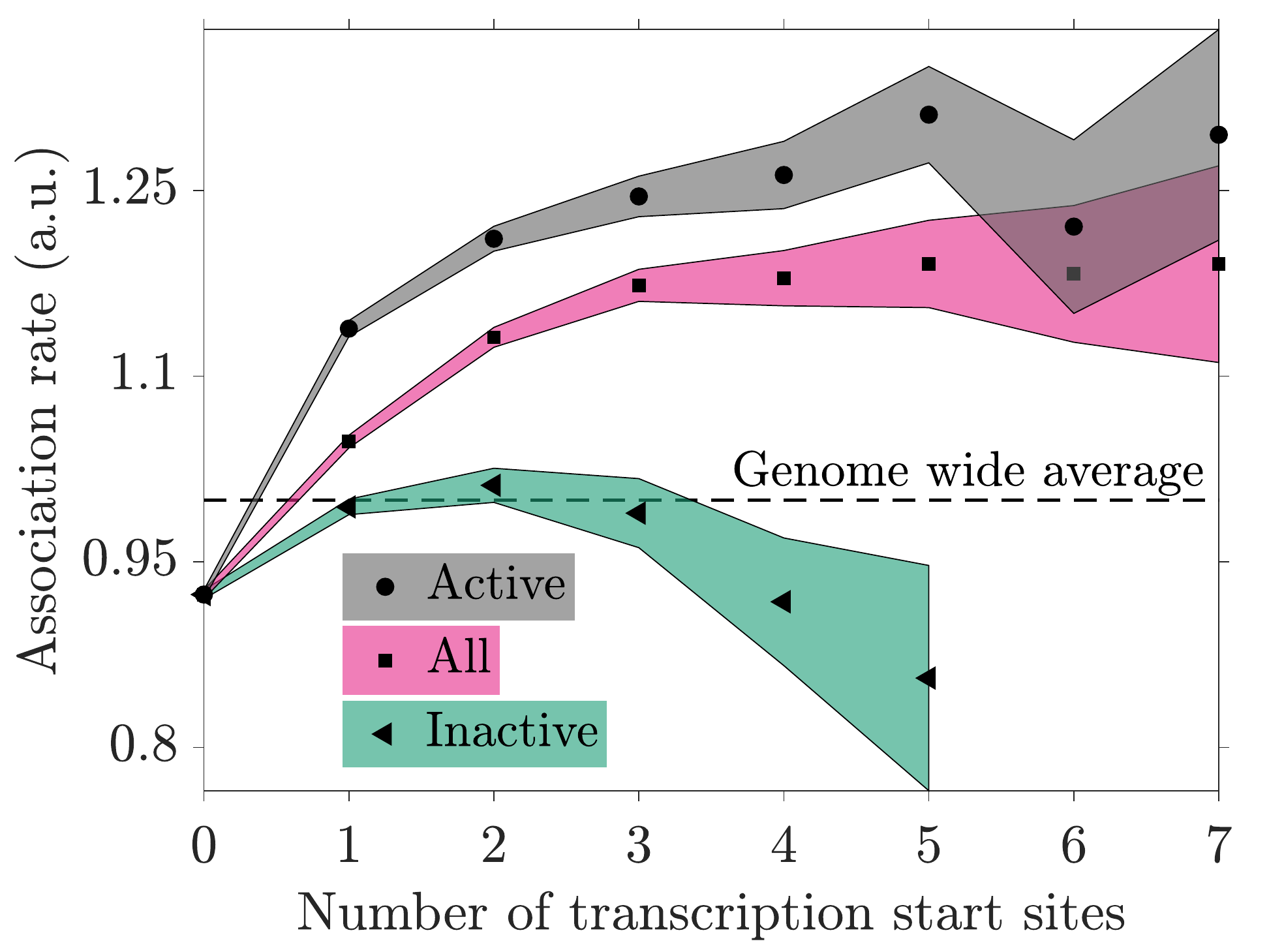}
    \caption{Nodes with many gene starts have higher predicted association rates -- and thus easier to find --  than nodes with few gene starts (in humans). We define the gene starts as the Transcription Start Sites (TSSs). The curves represent predicted association rates to nodes with active TSSs (gray), inactive TSSs (green), and any TSS type (pink). The active TSSs have higher association rates than the genome wide average ($\langle K_a \rangle=1$, dashed), whereas nodes with inactive TSSs (green) are below (apart from one data point).  The symbols represent the average association rate [Eq. \eqref{eq:fast+Ka}] and the colored areas show the 95\% confidence interval.  Parameters (dimensionless, see \cite{suppl}): $\koff=0.002$, $\kon=0.001$ and $\rho_0=0.5$. We omitted data points with less than 7 TSSs per node because the sample size is too small ($\leq$ 50 nodes).}
    \label{fig:mfptVSnTSS}
\end{figure}

\subsection{Chromatin regions with high association rates are enriched with active genes}

Figure \ref{fig:Ka_vs_chr} shows that the association rate varies across the genome. This is important  for regulatory proteins, such as transcription factors, that look for  promoters to control transcription. We therefore ask: are promoter regions easier to find than non-promoter regions? 

To answer this, we downloaded gene annotation data for human cells \cite{data} to extract the gene starts -- defined as the Transcription Start Sites (TSSs) -- and correlated them with the predicted association rates from Fig.~\ref{fig:Ka_vs_chr}. We found that the rates grow with the number of gene starts per node (Fig. \ref{fig:mfptVSnTSS}, pink). The data points represent the average association rate to all nodes with the same number of gene starts and the shaded area shows the 95\% confidence interval. In other words: gene-dense regions are easy to find.

Then we asked: because these regions harbour both active and inactive genes, are active gene-dense regions easier to find than inactive ones? To see this, we grouped the gene starts into  'active' and 'inactive' based on the RNA expression level surrounding each TSS -- 1kb upstream and downstream -- and calculated the association rates to the nodes with these TSSs. We found that nodes with many active TSSs have even higher association rates than if we do not separate active and inactive TSSs: gray is above pink in Fig. \ref{fig:mfptVSnTSS}. 

For nodes with inactive gene starts  we find the inverse relationship: green is below pink in Fig. \ref{fig:mfptVSnTSS}. This is underscored when comparing the green area to the genome-wide average $\langle K_a \rangle=1$  represented by dashed line: most inactive TSSs are in regions with association rates that are smaller than unity. This suggests that inactive gene starts are hard to find.

Figure \ref{fig:mfptVSnTSS} also shows that the association rate grows slowly beyond one or two TSSs per node: adding a few extra TSSs does not make the node easier to find. However, there is still a positive correlation between the number of gene starts per node and high association rates.

\subsection{Chromatin regions with  high RNA expression levels have high association rates}
Figure \ref{fig:mfptVSnTSS} suggests that transcription factors quickly find highly transcribed genes because the association rate is larger for active than inactive TSSs. But what about any transcribed region? Are the association rates high also for them?

\begin{figure}
\includegraphics[width=\columnwidth]{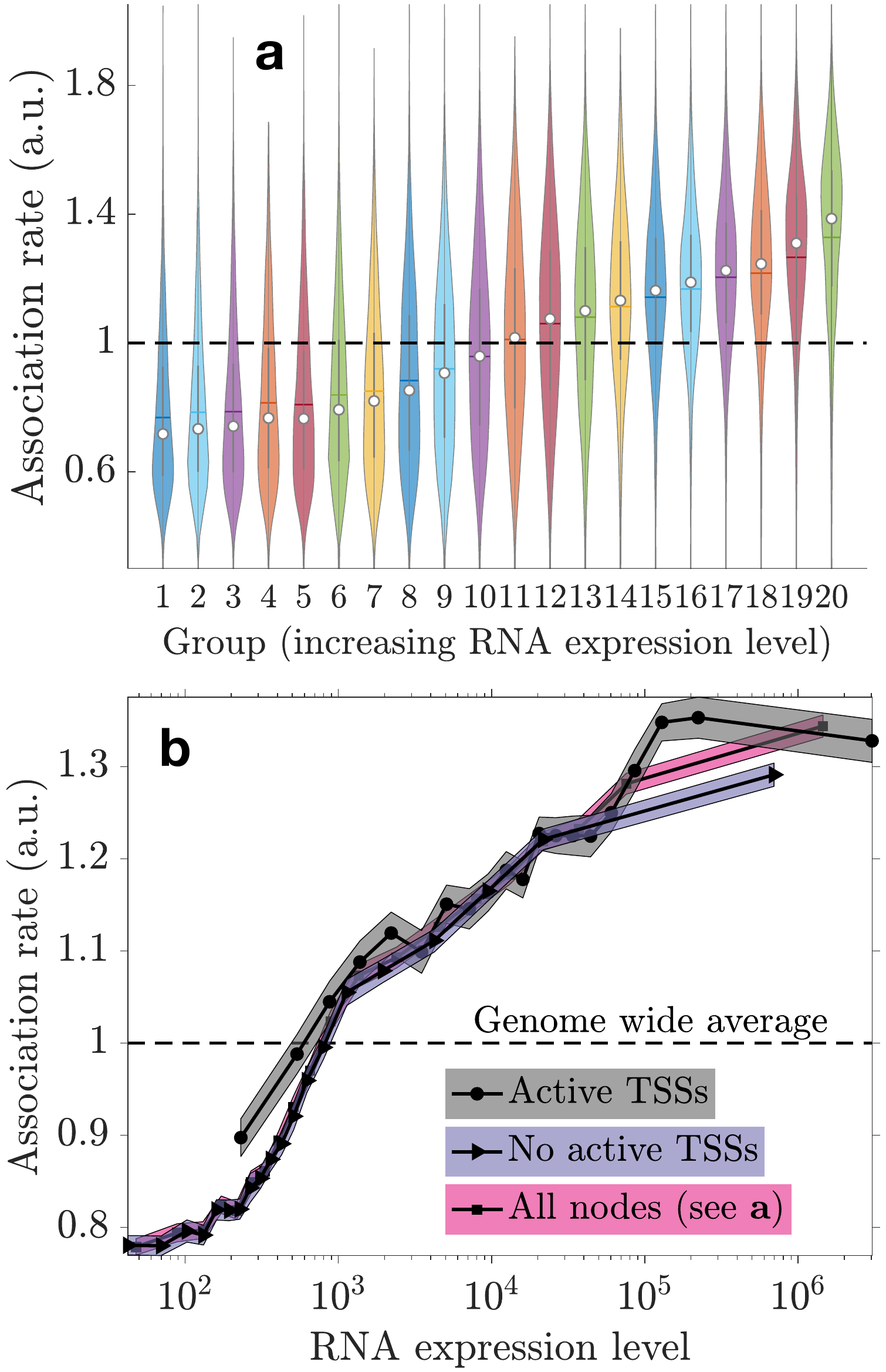}
\caption{Nodes that are highly transcribed are easy to find.
%
(a) The distribution of $K_a$ values for all nodes across the gnome. The nodes are ranked by their RNA expression level and divided into 20 groups. White circles represent the median and horizontal bars represents the mean. 10 groups lie above the genome wide average  ($\langle K_a \rangle=1$, dashed line). 
%
(b) Predicted association rate as function of RNA expression level for nodes with at least one active TSS (grey) and no active TSSs (purple). 
The RNA expression is divided into 20 bins with equally many points in each bin (same as in (a)).
%
The nodes with active TSSs tend to be above the genome wide average (18 points is above $\langle K_a \rangle=1$), while most nodes with no active TSSs are below (6 points above $\langle K_a \rangle=1$). The shaded areas show the 95\% confidence interval. Parameters used in both plots: $\koff=0.002$, $\kon=0.001$ and $\rho_0=0.5$.
\label{fig:mfptVSRNA}}
\end{figure}

To study this, we summed the RNA expression in all nodes across the genome and ranked them based on their RNA expression level. Then we partitioned the nodes into 20 equally-sized groups and calculated the association rate in each group. Shown as a violin plot (Fig. \ref{fig:mfptVSRNA}a),  we find that our predicted rates vary widely but that the median (white  circles) increases with high RNA expression levels (Spearman's correlation coefficient = 0.5449 \cite{spearman1904proof}). This suggests that nodes with high RNA expression levels -- with or without active gene starts --  are relatively easy to find.



To see by how much this correlation is caused by active gene starts, we made two new groups: nodes with at least one active TSS  and the rest -- nodes with inactive or no TSSs. Then, as before, we ranked the nodes in these large groups based on the RNA expression levels, divided them into 20  equally-sized subgroups, and calculated the average association rate for each subgroup. Plotting the predicted average association rate for the two large groups versus the average RNA expression level as well as the average for all nodes (Fig. \ref{fig:mfptVSRNA}a), we cannot see any significant difference: all curves  nearly lie on top of each other (Fig. \ref{fig:mfptVSRNA}b). 
This is a more general result than before. It is not only the highly transcribed gene starts that are relatively easy to find, it is any region with high RNA expression. 

In addition, as a simple measure of DNA accessibility,  we  checked how the association rate change with the fraction of base pairs that are transcribed per node \cite{suppl}. Just like for the RNA expression level, we find a positive correlation with high association rates (Spearman's correlation coefficient = 0.5632).

\section{Discussion and Summary}

Protein-binding experiments show that association rates change if the binding sites are embedded in a short or a long piece of DNA \cite{riggs1970lac}.  This is partly explained by the facilitated-diffusion model in which proteins switch between 1D search along the DNA, and 3D search -- by diffusion -- in the surrounding volume. The association rates also change if the DNA is straight or coiled \cite{van2008dna}. This can be understood if the facilitated diffusion model includes inter-segmental transfer between loop anchors \cite{lomholt2009facilitated}. However, current studies use standard polymer models that do not capture the chromatin's complex 3D organization in eukaryotes. To remedy this, we used Hi-C data as proxy for the  3D proximity between chromatin segments {\it in vivo}. This allowed us to to map the protein search problem onto a network problem with nodes and links representing chromatin segments and how they are physically connected to each other. Then we formulated a master equation for the number of searching proteins per node, from which we calculated analytically the genome-wide association rates in terms of the eigenvalues and eigenvectors of the Hi-C matrix. Using human Hi-C data, we compared the predicted association rates with RNA expression data and positions of gene starts. We found that regions which are easy to find -- measured by high association rates --  are enriched with active genes and have a generally high level of RNA expression.

We assume that the protein finds the target, for example a promoter site, as soon as it arrives at the target node -- here, 40 kb. This means that we model protein binding as diffusion-limited. However, some transcription factors, such as TetR in mammals \cite{normanno2015probing}, are suggested to be reaction-limited.  To include imperfect  protein-DNA binding our model we may follow \cite{berg2016helical}. Denoting the protein-DNA binding  rate as $k_{\rm DNA}$, and reinterpreting the on rate $\kon$ as an effective on rate $\kon^{\rm eff.}$, we may write  $1/\kon^{\rm eff.} = 1/\kon + 1/k_{\rm DNA}$ where  $K_a = \kon^{\rm eff.} + \gamma_a V_a$.


We calculated the association rates in each chromosome without considering inter-chromosome connections. This is an assumption as chromosomes  do come in physical contact. From Hi-C data, however, it seems like these contacts are less frequent than within chromosomes.  This is a limitation of the data rather than our model that  can handle any genome-wide Hi-C map.




Overall, this study provides a framework to predict protein-binding positions dictated by chromatin contact maps in the cell nucleus. As such, it opens new ways to interpret binding profiles of transcription factors that cannot be explained by the DNA sequence \cite{farnham2009insights,schmidt2010five}. Mechanistic understanding of these profiles is important to reach a molecular understanding of  gene regulation.

\section{acknowledgements}
TA and LL acknowledges support from Swedish Research Council (grant no: 2014-4305 and 2017-03848). PS acknowledges support from the Knut and Alice Wallenberg foundation (grant no. 2014–0018, to EpiCoN, co-PI: PS). We thank Dr. Rajendra Kumar for his help with gcMapExplorer  \cite{kumar2017genome} to analyze the Hi-C data.

\bibliographystyle{unsrt}
\bibliography{refs}

\newpage~
\newpage
\appendix
\section{SUPPLEMENTARY MATERIAL}

\section{Particle flux through the target}
The number of proteins that reached the target up to time $t$ is $J_a(t)$. For non-zero $\kon$ and $\koff$ it reads
\be \label{eq:Ja+full}
\begin{aligned}
J_a(t) &= \sum_j \omega_{aj}\sum_i V_{ji} \left\{ \frac{\kon^i}{\koff-\lambda_i}\left[ t-\frac{1-e^{-t(\koff-\lambda_i)}}{\koff-\lambda_i} \right] \right. \\  & + \left. \frac{1-e^{-t(\koff-\lambda_i)}}{\koff-\lambda_i}\rho_0\sum_{l\neq a}V^{-1}_{il} \right\} +\kon t.
\end{aligned}
\ee
where  $\lambda_j$ and  $V_{ij}$ are the eigenvalues and of eigenvectors $\omega_{ij}$. Because $\lambda_1=0$ is the largest eigenvalue, we can approximate Eq. \eqref{eq:Ja+full} at times $t\gg \koff^{-1}$ with terms proportional to $t$ 
\be \label{eq:J+approx}
J_a(t) \simeq \left(\kon + T_a^{-1} \right)t,\quad T_a^{-1}=\sum_j\omega_{aj}\bar{n}_j,
\ee
where the steady-state distribution is
\be \label{eq:steady_state_exact}
\bar{n}_j=\sum_i\frac{V_{ji}\kon^{i}}{\koff-\lambda_i},
\ee

The relation $J_a(t)\simeq (\kon+T_a^{-1}) t$ coincides with the continuum approach in \cite{lomholt2005optimal} for proteins that combines bulk excursions with 1D sliding (jumping to nearest neighbours in our model) and L\'evy relocation's with jump lengths $x$ distributed like $\simeq|x|^{-1-\alpha}$ $(0<\alpha<2)$. Since $\omega_{ij}\simeq|i-j|^{-1-\alpha}$ with $0<\alpha<1$ -- on average -- we see that our model is a network analogue of t \cite{lomholt2005optimal}.

\section{Particle flux through the target without bulk exchange}
\label{app:k=0}
Here we investigate the case when proteins do not unbind from  the DNA. As $\kon, \koff \to 0$, Eq. \eqref{eq:Ja+full} becomes
\be \label{eq:J_binding_zero}
\begin{aligned}
J_a(t)&=\sum_{k=1}^N\omega_{ak}  \sum_{i=2}^N\frac{V_{ki}}{|\lambda_i|}(1-e^{-t|\lambda_i|})\sum_{j\neq a}\rho_0V^{-1}_{ij}\\
&=N_p- \sum_{k=1}^N\omega_{ak}  \sum_{i=2}^N\frac{V_{ki}}{|\lambda_i|}e^{-t|\lambda_i|}\sum_{j\neq a}\rho_0V^{-1}_{ij},
\end{aligned}
\ee
with $N_p=\rho_0(N-1)$. For large times we know that $J_a(t\to\infty)=N_p$ since by then all proteins have arrived to the target. This leads to the simplification in the 2nd row.  For large times $t\ll |\lambda_N|^{-1}$ --  $\lambda_N$ is the largest eigenvalue (in magnitude) -- where $J_a(t)\ll N_p$, we find the same behaviour as before, $J_a(t)\propto t$. This  is seen by expanding Eq. \eqref{eq:J_binding_zero} around $t=0$. 

\section{Derivation of Eq. (6) in the manuscript} \label{sec:approx}
\subsection{Fast target finding ($\koff \ll 1$)}
When the unbinding rate $\koff$ is small compared to the association rate $K_a$, the number of proteins per node is close to its initial value $\rho_0$ by the time of the first arrival to the target, and we have the approximation
\be \label{eq:approx_fast}
K_a = \kon + \rho_0 W_a,
\ee
where $W_a=\sum_{i\neq a}\omega_{ai}$. We may find this approximation by expanding Eq. \eqref{eq:Ja+full} around $t=0$ and using the inverse transformation $\sum_jV_{ij}q_j(0)=n_i(0)=\rho_0(1-\delta_{ia})$. 

Furthermore, by demanding that the genome wide average of $K_a$ in Eq. \eqref{eq:approx_fast} is unity, $\langle K_a \rangle = 1$, and using that $\omega_{ai}=\fcoll v_{ai}$ we find $\fcoll$ to be
\be \label{eq:fcoll+unity}
\fcoll = \frac{1-\kon}{\rho_0 \langle V_a \rangle}.
\ee
Using this in Eq. \eqref{eq:approx_fast} leads to
\be \label{eq:approx_fast_fcoll}
K_a = \kon + (1-\kon)\frac{V_a}{\langle V_a \rangle}, \ \ \ \ \koff \ll 1.
\ee
With this definition of $\fcoll$, the binding rate $\kon$ is bound by $[0,1]$, where  $\kon=1$ corresponds to target-finding directly from the bulk. 

\subsection{Target finding in steady-state ($\koff \gg 1$)}
When the unbinding rate $\koff$ is large compared to the association rate $K_a$, few proteins will find the target before leaving on a bulk excursion. In this limit, the system reaches its steady-state --  with $\bar{\rho}_a$ the number of proteins per node --  before the first arrival to the target. This leads to the approximation
\be \label{eq:Ka_slow_approx}
K_a = \kon + \bar{\rho}_aW_a.
\ee

To arrive at this equation we identify in  Eq. \eqref{eq:J+approx} that $J(t) \simeq K_at$. Then we replace the $\bar{n}_j$ by the approximate density $\bar{\rho}_a$ that we find $\bar{\rho}_a$ by the following argument. In  steady state, proteins bind to  the DNA with rate $\kon$. Except for the absorbing target, there are $N-1$ nodes available to bind. Similarly, there are $\bar{\rho}_a(N-1)$ number of proteins that unbind from the DNA with rate $\koff$. Last, proteins are absorbed at the target with rate $T_a^{-1}=\bar{\rho}_aW_a$. These three terms sum to zero, and therefore
\be 
\bar{\rho}_a=\frac{\kon}{\koff+W_a/(N-1)}.
\ee
Using that $W_a = \fcoll V_a$ with $\fcoll$ from Eq. \eqref{eq:fcoll+unity} gives
\be \label{eq:Ka_slow_scale}
K_a = \kon + \frac{\kon(1-\kon)}{\kon + \frac{1-\kon}{N-1}\frac{V_a}{\langle V_a \rangle}}\frac{V_a}{\langle V_a \rangle}, 
\ \ \ \ \koff \gg 1.
\ee
%

\section{Validation of approximations} \label{sec:approx_valid}
\begin{figure}
\includegraphics[width=1\columnwidth]{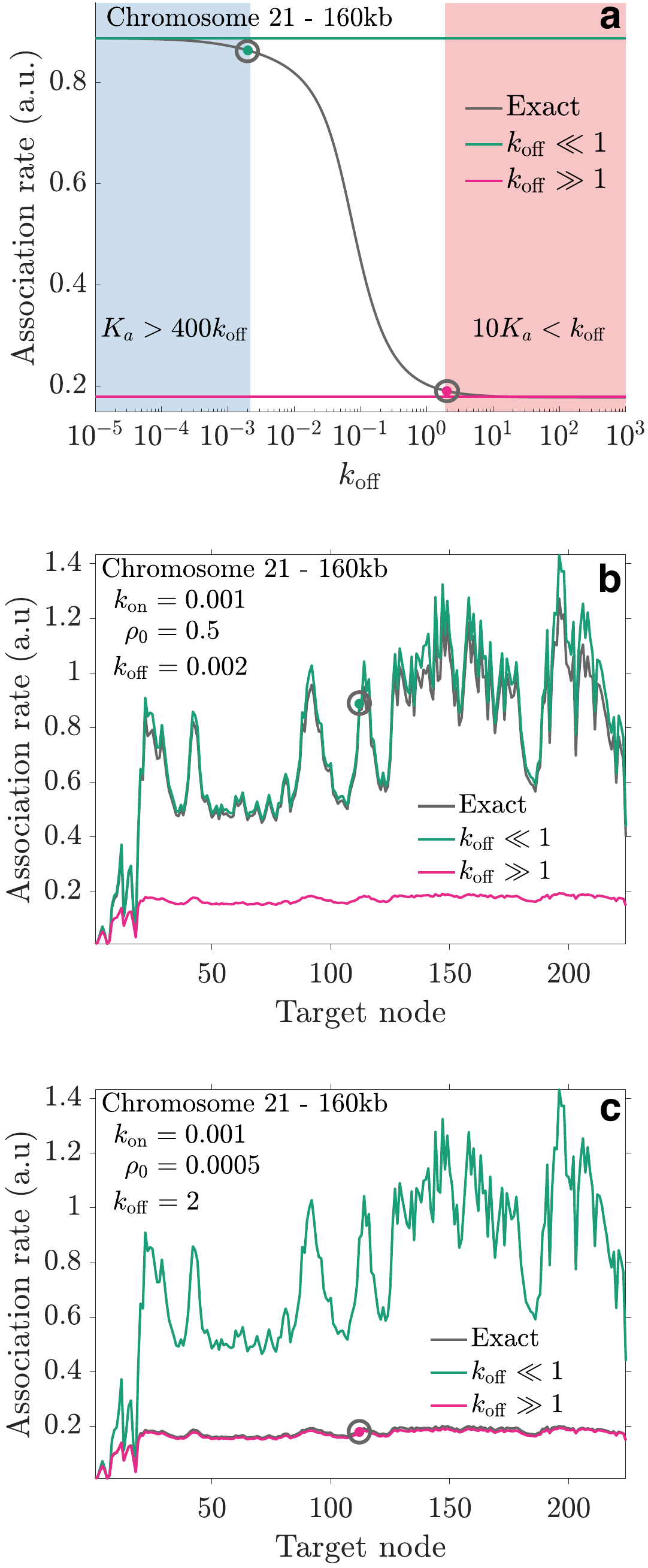}
\caption{Comparison of the approximations to the exact, time-integrated association rate, using chromosome 21 at the resolution 160 kb. {\helv{a}} In the blue area where $\koff\ll1$ the target finding is fast, while in the red area where $\koff\gg1$, the system is close to its steady state and target finding is slower. Target at $N/2$, in the middle of  the system. {\helv{b}} Association rates to all targets when $\koff\ll1$. {\helv{c}} Association rates to all targets when $\koff\gg1$. Note the green and red circled dots in {\helv{b}} and {\helv{c}}, they correspond to the same parameter values as in {\helv{a}}, respectively.\label{fig:france}}
\end{figure}
To better understand the validity of Eqs. \eqref{eq:approx_fast_fcoll} and \eqref{eq:Ka_slow_scale}, we compare them to  the exact association rate %
\begin{equation} \label{eq:Ka_exact}
K_a^{\rm exact}=\left(\int_0^{\infty}\exp(-J_a(t))dt\right)^{-1}. 
\end{equation}

Figure \ref{fig:france}{\helv{a}} shows how the association rate changes for a specific target node -- we choose $a=N/2$ in human chromosome 21 -- as we change $\koff$ while keeping the on-rate fixed, $\kon=0.001$, and adjusting the density $\rho_0=\kon/\koff$. The solid grey line shows $K_a^{\rm exact}$  and the horizontal lines represent the approximations for small and large $\koff$ -- Eqs. \eqref{eq:approx_fast_fcoll} and \eqref{eq:Ka_slow_scale}.

The blue area in Fig. \ref{fig:france}{\helv{a}} shows the large-$K_a$ regime ($K_a>400\koff$). Here,  Eq. \eqref{eq:approx_fast_fcoll} deviates only a few percent from $K_a^{\rm exact}$: the deviation is $2.7\% (\approx 1- K_a/K_a^{\rm exact})$ at the encircled green dot. To get this number,  we used $\koff=0.002$, $\kon=0.001$ and $\rho_0=0.5$ --  the same values that we used to create all plots in the main text. 

Th pink area represents the opposite limit: small $K_a$ ($K_a<\koff/10$). In this region the approximation Eq. \eqref{eq:Ka_slow_scale} is a good match to  $K_a^{\rm exact}$. At the red dot ($\koff = 2$) the relative error  is 5.7\%.

In the intermediate region (white area), we cannot use the simple expressions because the flux $J(t)$ has a complicated time-dependence. To get the association rate in this regime, we have to evaluate Eq. \eqref{eq:Ka_exact} directly.

In Figs. \ref{fig:france}{\helv{b}} and \ref{fig:france}{\helv{c}}, we calculate the association rate for all nodes in chromosome 21 using Eqs. \eqref{eq:Ka_exact}, \eqref{eq:approx_fast_fcoll} and \eqref{eq:Ka_slow_scale}  with fixed parameters (shown in the figures).  The figure shows the limiting $\koff$ cases. In \ref{fig:france}{\helv{b}} the unbinding rate is small ($\koff=0.002$), and we see that the approximation \eqref{eq:approx_fast_fcoll}  match well with $K_a^{\rm exact}$ whereas Eq. \eqref{eq:Ka_slow_scale}  does not. Equation \eqref{eq:Ka_slow_scale} that matches better in \ref{fig:france}{\helv{c}}  where the unbinding rate is large ($\koff=2$).

\begin{figure*}
\includegraphics[width = \textwidth]{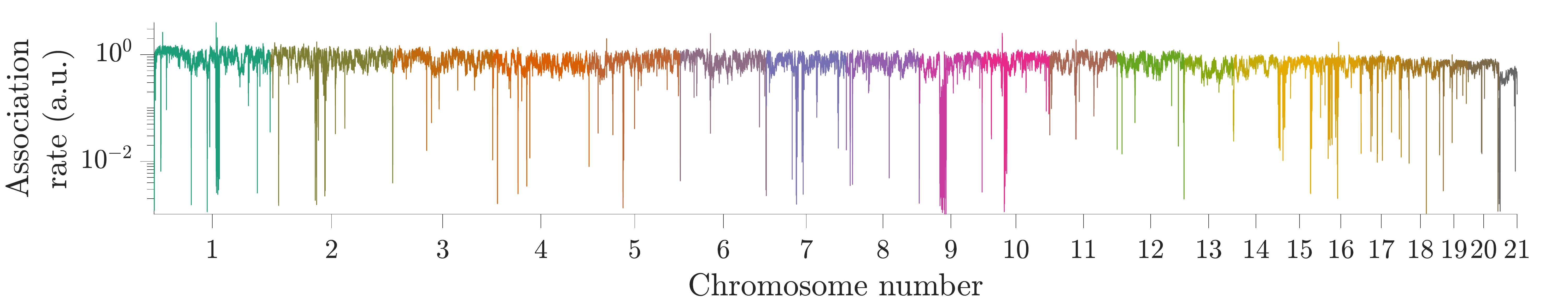}
\caption{Genome-wide association rates $K_a$ at 40 kb resolution evaluated using Equation \eqref{eq:Ka_slow_scale}.  The association rates vary by several orders of magnitude along the chromosomes but the 95\% confidence interval of $K_a$ varies only by a few percent ($0.7531\pm0.0018$). We used these parameters: $\koff = 2$, $\kon = 0.001$ and $\rho_0=0.0005$.}
\label{fig:Ka_vs_chr_slow}
\end{figure*}
\section{Genome-wide association rates when $\koff\gg 1 $} \label{sec:Ka_vs_chr_slow}
Similar to Fig. 2 in the main text, we  show the association rate for all targets in every chromosome at large $\koff$. We calculate $K_a$ from Eq.  \eqref{eq:Ka_slow_scale}, see Fig. \ref{fig:Ka_vs_chr_slow}. These curves for $\koff\gg 1$ and $\koff\ll 1$ (main text) are almost identical, except by an offset on the $y$-axis.

\section{Genome-wide association rate as a function of node strength} \label{sec:Ka_vs_Va}

\begin{figure}
\includegraphics[width = \columnwidth]{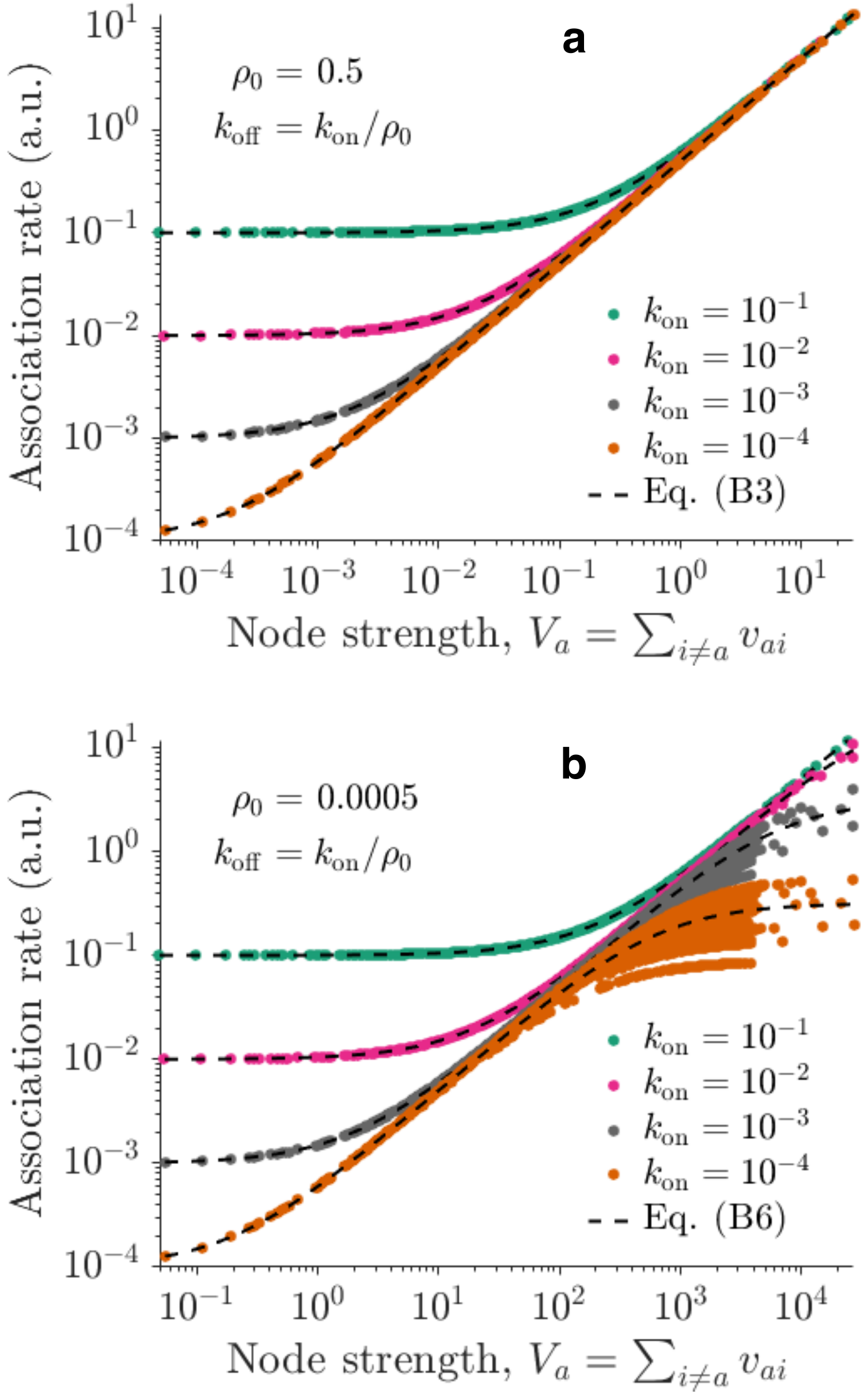}
\caption{The association rate vs the nodes' strength $V_a$ (sum of all link weights). {\helv{a}} All data points follow the universal law $K_a = \kon + (1-\kon)V_a/\langle V_a \rangle$, see Eq. \eqref{eq:approx_fast_fcoll}. The density $\rho_0=0.5$ is kept fixed in all four cases as we increase $\kon$. {\helv{b}} Association rate in  during steady state ($\koff\gg1$) calculated from Eq.\eqref{eq:Ka_slow_scale}. The behavior is not universal as it depends on the size $N$ of each chromosome (number of nodes). The dotted line evaluated analytically -- as $N$ we picked the mean chromosome size.} \label{fig:Ka_vs_Va_fast}
\end{figure}

In Fig. \ref{fig:Ka_vs_Va_fast}{\helv{a}} we show $K_a$ -- calculated from Eq. \eqref{eq:approx_fast_fcoll}  ($\koff \ll 1 $) -- varies with node strength $V_a$ for our different values of $\kon$ with fixed  $\rho_0=0.5$all nodes across the genome; The symbols represent values for individual nodes across the human genome. For comparison we  plot the analytical prediction Eq. \eqref{eq:approx_fast_fcoll}. We find that the search times are dominated by $\kon$ for weakly connected nodes. For strongly connected ndoes, we find the universal behavior $K_a\propto V_a$. 

In Fig. \ref{fig:Ka_vs_Va_fast}{\helv{b}} we show  $K_a$ for all nodes in the other limit  $\koff \gg 1 $. Here $K_a$ depends on the number of nodes $N$ -- via $\rho_a$ in Eq. \eqref{eq:Ka_slow_scale} -- and therefore we do not expect a universal large-$V_a$  behaviour 

\section{Fraction of transcribed DNA} \label{sec:frac_DNA}
\begin{figure}
    \centering
    \includegraphics[width=\columnwidth]{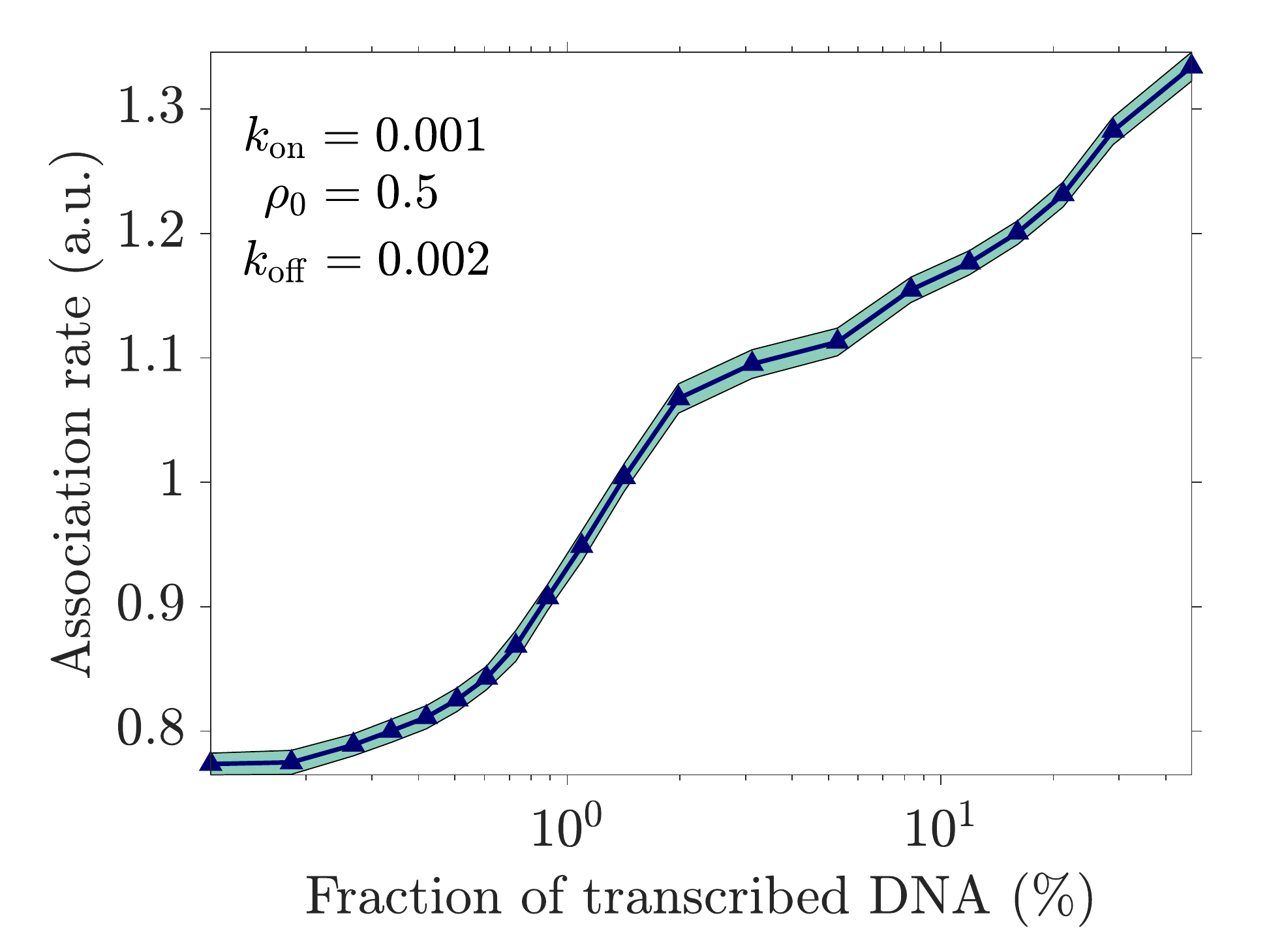}
    \caption{Association rate increases with the fraction of transcribed DNA. This is similar to the RNA expression level which also increases with the association rate.}
    \label{fig:Ka_vs_openChr}
\end{figure}

We use RNA expression data (downloaded from ENCODE) to calculate the fraction of transcribed DNA. This is calculated in the following way. For every 40 kb region $i$ across the genome, we count the number of base pairs $n_i$ that has an RNA expression level above zero. The fraction of transcribed DNA in region $i$ is thus $n_i/40000$. In Fig. \ref{fig:Ka_vs_openChr} we plot the association rate as a function of the fraction of transcribed DNA, where the all nodes are divided into 20 equally sized groups. The correlation between $K_a$ and fraction of transcribed DNA is slightly stronger than for RNA expression level (Spearman correlation coefficient = 0.5632). We point out that the fraction of transcribed DNA and the RNA expression correlate strongly: Spearman's correlation coefficient is 0.9693.
\begin{figure}
    \centering
    \includegraphics[width=\columnwidth]{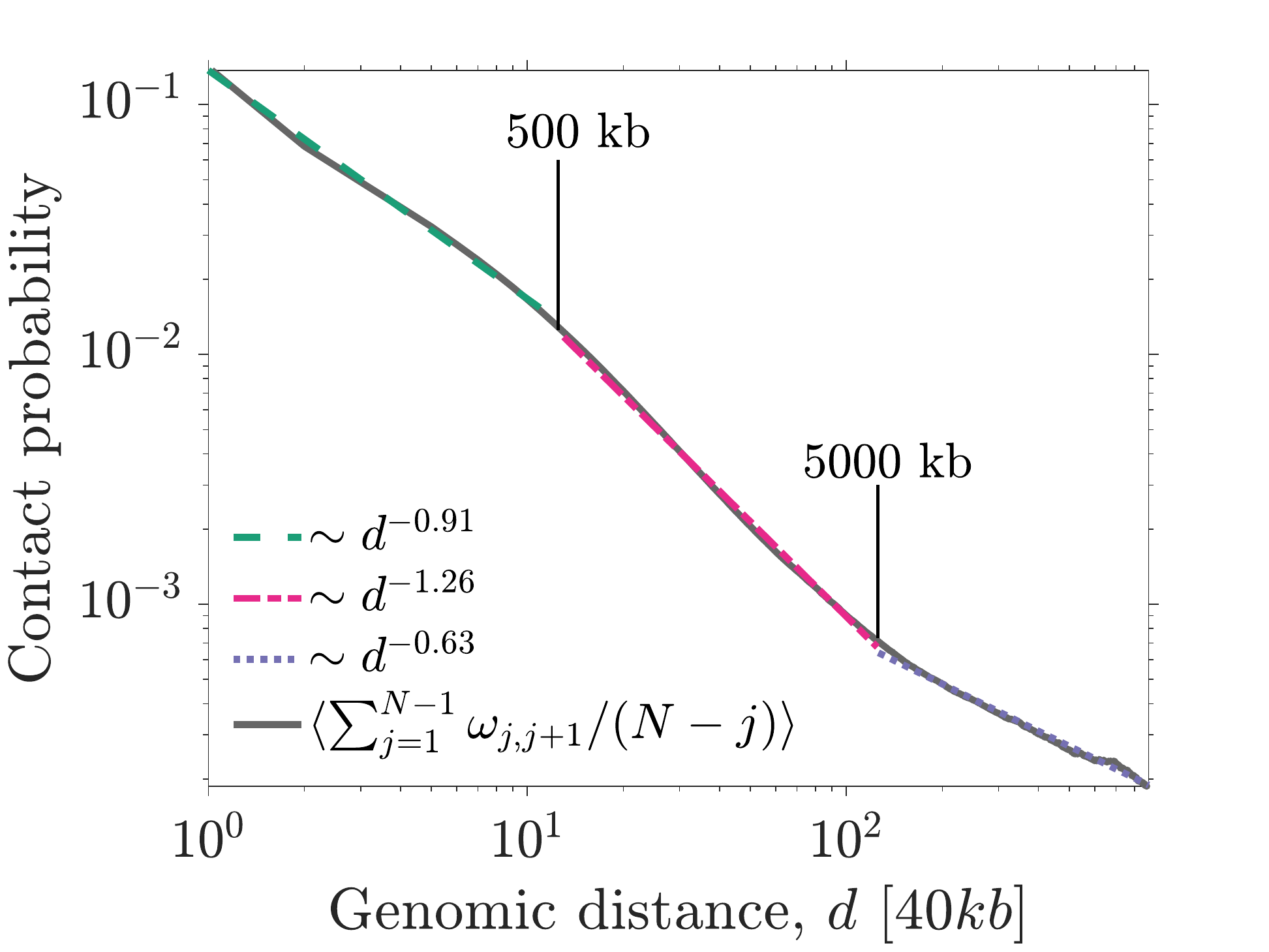}
    \caption{The average contact probability (genome wide) decays as $d^{-\alpha}$  with distance $d$ from the diagonal. At   $d=500$ kb and $d=5000$ kb, $\alpha$ changes value. The black line is is the averaged data from $\omega_{ij}$, and green, pink and purple are   fits.}
    \label{fig:contact_prob}
\end{figure}

\section{Contact probability} \label{sec:contact_prob}
The average contact probability decays with distance from the diagonal. To see this, we calculated  $p_{ij}\ = \langle \sum_{j=1}^{N-1}\omega_{j,j+1}/(N-j)\rangle$, where $\langle ... \rangle$ denotes genome-wide average. We find that $p_{ij} \sim |i-j|^{-\alpha}$ where there are  three regimes with different $\alpha$ (Fig. \ref{fig:contact_prob}). Since the chromosomes has different sizes, the regimes appear at different length scales, but  the cross-overs are roughly at 500 kb and 5,000 kb. 
We used Hi-C matrices $\omega_{ij}=\fcoll V_{ai}$, where $\fcoll$ is given in equation \eqref{eq:fcoll+unity} with $\kon=0.001$ and $\rho_0=0.5$.

\section{Transcription start sites (TSS\lowercase{s})}\label{how_to_TSS}
To distinguish between active and inactive TSSs we use  RNA expression data (downloaded from ENCODE) and calculated the average number of RNA reads per base pair, $\bar n_{\rm RNA}$, 1kb upstream and 1kb downstream of each TSS. We defined an active TSS as when $\bar n_{\rm RNA}\geq 1$. Given this threshold we found 32712 active and 20795 inactive TSSs.

\begin{figure*}[]
\includegraphics[width=0.79\textwidth]{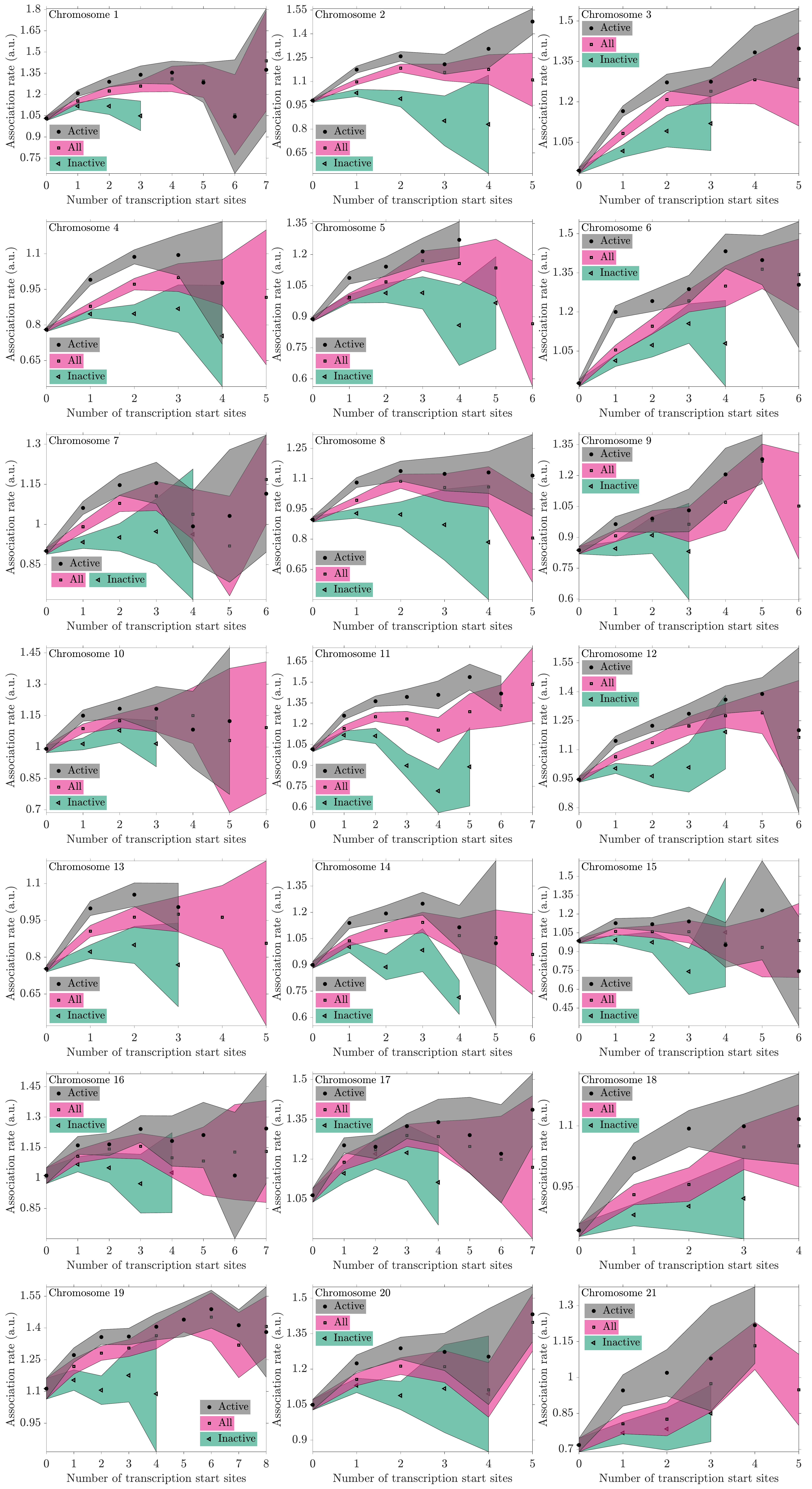}
\caption{The association rate $K_a$ changes with locations of gene starts and RNA transcription levels for each chromosome. We truncated the plots where it was less than 7 $K_a$ values to calculate the average association rate. Parameter values: $\koff=0.002$ $\kon=0.001$ and $\rho_0=0.5$. $K_a$ evaluated using Eq. \eqref{eq:approx_fast_fcoll}}.
\label{fig:all+TSS}
\end{figure*}

For each chromosome at 40 kb resolution, we show  howÊ$K_a$ changes with the number of TSSs per node (Fig. \ref{fig:all+TSS}). The plots show three TSS groups:  active,  inactive , and all. The genome-wide average of all these plots is Fig. 3 in the main text.

\bibliographystyle{unsrt}
\bibliography{refs}

\end{document}